\documentclass[aps,twocolumn,letterpaper,prl,showpacs]{revtex4}

\usepackage{amsfonts}
\usepackage{amsmath}
\usepackage{amssymb}
\usepackage{graphicx}

\makeatletter
\def\lsim{\mathrel{\mathpalette\@versim<}}
\def\gsim{\mathrel{\mathpalette\@versim>}}
\def\@versim#1#2{\vcenter{\offinterlineskip
	\ialign{$\m@th#1\hfil##\hfil$\crcr#2\crcr\sim\crcr } }}
\makeatother

\begin{document}

\title{Brush-like Interactions between Thermoresponsive Microgel Particles}
\author{Frank Scheffold$^{(1,\ast)}$, Pedro D\'{\i}az-Leyva$^{(1)}$, Mathias Reufer$^{(1,2)}$, Nasser Ben Braham $^{(1)}$, Iseult Lynch$^{(3)}$ and James L. Harden$^{(4)}$  }
\address{$^{(1)}$Dep.~of Physics and Fribourg Center for Nanomaterials,~Univ.~of~Fribourg,~1700~Fribourg,~Switzerland}
\address{$^{(2)}$Adolphe Merkle Institute, University of Fribourg, 1723 Marly, Switzerland}
\address{$^{(3)}$ School of Chemistry and Chemical Biology, University College Dublin, Belfield, Dublin 4, Ireland }
\address{$^{(4)}$ Department of Physics, University of Ottawa, Ottawa, Ontario K1N 6N5, Canada }

\date{\today}

\begin{abstract}
Using a simplified microstructural picture we show that interactions between thermosensitive microgel particles can be described by a polymer brushlike corona decorating the dense core.  The softness of the potential is set by the relative thickness $L_0$ of the compliant corona with respect to the overall size of the swollen particle $R$. The elastic modulus in quenched solid phases derived from the potential is found to be in excellent agreement with diffusing wave spectroscopy data and mechanical rheometry. Our model thus provides design rules for the microgel architecture and opens a route to tailor rheological properties of pasty materials.
\end{abstract}


\maketitle
Thermoresponsive microgel particles are a hybrid between a colloidal and a polymeric system with properties that can be tuned externally ~\cite{Microgel1,Microgel2,Stieger2004}. Most  of the previously studied stimuli responsive microgel systems are based on poly(N-isopropyl-acrylamide) (PNIPAM), a polymer which has a
lower critical solution temperature (LCST) of approximately $33^{\circ}$C ~\cite{Microgel1}.
Above the LCST, the microgel particles expel water and are collapsed.  The typical size of a collapsed microgel particle is in the range of 0.2-1 $\mu$m. Upon lowering the temperature the particles swell to about twice their original size.  Responsive microgels thus provide the possibility to fabricate \emph{smart} colloidal materials for applications as viscosity modifiers, carrier systems, bioseparators, optical switches or sensors ~\cite{Microgel1,Microgel2,Weissman96}. Moreover, due to their tunability they are ideal model systems to study the phase behavior, glass transition and jamming in dense colloidal dispersions \cite{Lyon2007,JWu2003}.
\newline \indent Particles come into contact and form a viscoelastic paste below the LCST if the polymer density inside the swollen particles approaches the total polymer density. A number of rheological studies have revealed the apparent divergence of the viscosity at the transition point
and the emergence of an elastic shear modulus ~\cite{Microgel2, Stieger2004}. 
Since the particles consist of polymers and solvent in equilibrium they are elastically compliant, 
and thus elastic moduli do not diverge at the jamming transition, unlike the behavior of rigid colloidal particles \cite{Liu98,Brady93}. Several experimental studies show that in this regime the bulk modulus as a function of the effective volume fraction scales as a power law 
$G_p \propto \phi_{eff}^{1+n/3}$ with  values of n ranging from $n=9$ to $n=22$  \cite{Microgel2,Stieger2004}. 
These findings indicate that the particle interaction potential $\psi(r) \propto r^{-n}$ also follows a power law, 
at least over some significant range of length scales \cite{Microgel2,Stieger2004}. Unfortunately a  detailed model for the origin of these
interactions has been missing. This is mainly due to the fact that modeling the interaction between swollen particles is
complicated by the heterogeneous microstructure arising from the faster reaction rate of cross-linking 
compared to polymerization \cite{Wu94}. In this Letter we show that interactions between thermosenstive microgel particles can be modeled by a polymer brushlike corona decorating the dense core ~\cite{Alexander77,Witten86,Milner88}.  The softness of the potential is set by the relative thickness of the compliant corona with respect to the overall size of the swollen particle.
\begin{figure}[h!]
\includegraphics[trim=.3cm .5cm 4.5cm .5cm,clip,angle=270,width=7.5cm]{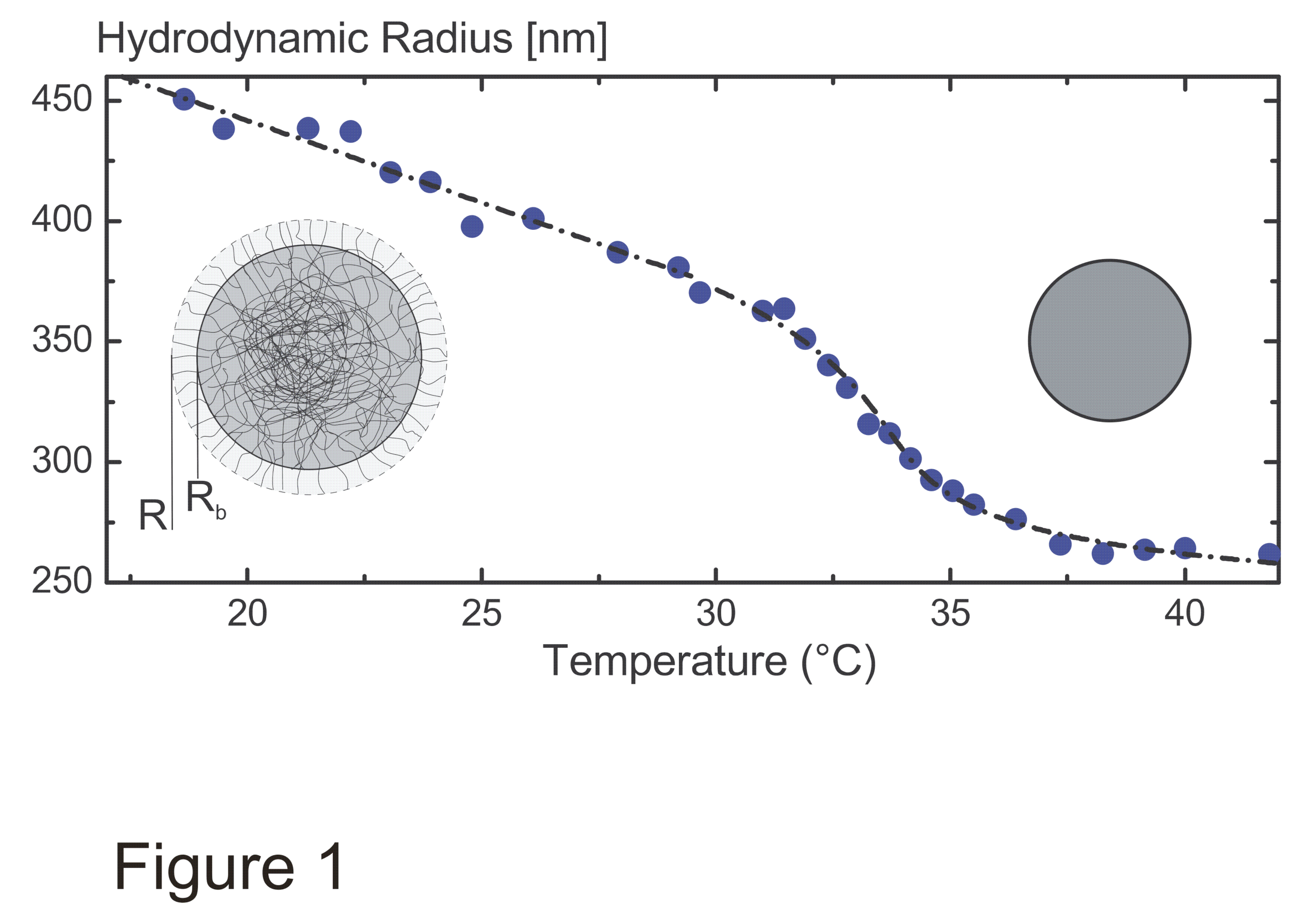}
\caption{PNIPAM hydrodynamic radius $R_H$ as obtained from dynamic light scattering. 
Dash-dotted line shows the interpolation curve used to calculated the effective volume fraction 
$\Phi_{eff} \propto R_H^3$. The sketch (drawn to scale) illustrates the swelling of the particles: 
Our experiments suggest that  the particles consist of a highly cross-linked core with radius $R_b$ 
and a corona of uncross-linked polymer chains  that can swell to a brush of thickness 
$L_0\simeq R-R_b$ under good solvent conditions (left).} 
\label{GraphicSinglePart} 
\end{figure}
Our PNIPAM microgel particles have been synthesized by standard methods and characterized 
as described in Ref. \cite{ReuferEPJE08}. The high cross-linker to monomer ratio of 5.3 mol $\%$ N,N'-methylene-bis-acrylamide (BIS) leads to relatively rigid microgel cores with, on average, one molecule of cross-linker 
per 19 molecules of NIPAM monomer. In Fig. \ref{GraphicSinglePart} we show the temperature dependence of the hydrodynamic radius $R_H$ determined by dynamic light scattering. The particle radius in the collapsed state is $R_0 \simeq 250$nm for which the internal polymer volume fraction is approximately $\phi_0 \approx 70\%$ \cite{ReuferEPJE08}.  For T$=20^{\circ}$C the particles reach a hydrodynamic radius of $R_H = 440$nm, 
almost doubling their size. As particles swell they acquire a radially inhomogeneous density profile. While the exact density distribution $\phi(\rho)$ in the swollen state remains unknown, scattering experiments clearly indicate the presence of a moderately swollen core and a highly swollen corona ~\cite{ANieves,Stieger2004,MasonPnipam,ReuferEPJE08}. For our particles we find from static light scattering a polymer density of $\phi\simeq 35 \%$ in the core and $\phi < 10 \%$ 
in the corona  (at $\rho > 0.8 R_H$) \cite{ReuferEPJE08}. 
Classical Flory-Rehner theory \cite{ANieves,FloryBook} for swollen polymer gels provides estimates for the degree of polymerization between cross-links $N_c$ in the core and corona regions.  
Using literature data for the statistical segment length $a \simeq 0.81$nm \cite{Kubota90} 
and the Flory interaction parameter $\chi$  \cite{Hirotsu2002} of NIPAM,  we find $N_c \sim 10$ in the core of the particle, 
in agreement with the stoichiometry of the microgel synthesis.  Moreover, for typical corona densities we find  $N_c \gg100$, suggesting an essentially uncrosslinked corona region for our polymerization conditions. 
\indent Below the LCST $\sim$ these corona chains extend into the good solvent and form a swollen brushlike layer of thickness $L_0$. While the transition between the core and corona regions is not an abrupt one, we adopt the simplified scaling picture of 
Alexander~\cite{Alexander77} of the brush as a collection of identical chains of length $N$ grafted at moderate-to-high 
surface density to a reference surface at $\rho=R_b$.   In good solvent conditions, the thickness of such a brush scales as 
$L_0\sim N s^{-2/3}a^{5/3}$, where $s \sim a \phi_b^{-3/4}$ is the average separation between grafting sites.   
Note that other more realistic models also predict $L_0\sim N$~\cite{Milner88}, so the simplicity of our model should not
adversely affect the results at the scaling level.  
Results from static light scattering indicate typical values of $\phi_b \sim 3 \%$ and $L_0 \sim$ 100nm for our particles \cite{ReuferEPJE08}. 
From these values we can estimate $s \sim 10$nm and $N\sim 700$. This means the corona density is well above the overlap concentration $\phi^* \simeq N^{-4/5}$ and $s < L_0$ which shows the consistency of our arguments. We note that due to the high degree of cross-linking the stiffness 
of the core region is substantially higher than the soft brushlike 
corona \cite{FloryBook}. 
More generally, as long as the number of monomers between cross-links is less
than the number of monomers in a corona chain, the core will be less compliant
than the corona region.  As we are well within this regime, we may treat the 
core as an effectively incompressible solid for the weak to moderate particle 
deformations considered. Here, the choice of a reference surface, $R_b$, 
is somewhat arbitrary but should reflect the boundary between core and corona behavior. 

\indent We now turn our attention to the properties of a dense microgel assembly with a total polymer concentration of about
$135$ mg/ml \cite{ReuferEPJE08}. From laser scanning
confocal microscopy studies of a diluted sample we determined the particle number density of the dense system to be
$\eta=2.24/\mu m^{3} \pm 2 \%$.  Over the range of temperatures covered, the effective particle volume fraction $\Phi_{eff}$ thus increases from about $0.2$ to $0.9$.  We use two-cell diffusing wave spectroscopy (DWS) in the transmission geometry to study particle diffusion, 
dynamical arrest, and build up of elasticity \cite{weitzpine}.   
DWS, dynamic light scattering in the highly multiple-scattering limit, provides information on the thermally-induced 
fluctuations of the microgel particles on the nanoscale via the temporal intensity autocorrelation function $g_2(t)$ of multiply scattered light.
In particular, the ensemble-averaged mean square displacement (MSD), $\left\langle {\Delta r^2 (t)} \right\rangle$, can be extracted
from $g_2(t)$. Details of the setup can be found in Ref. \cite{Zakharov06}.  
The sample is kept in a glass cuvette with inner dimensions $10 \times  2$ mm (Hellma, Germany). 
To avoid crystallization the sample is loaded hot, quenched below $20^{\circ}C$, and 
subsequently the temperature is increased in steps of $0.5^{\circ}C$ or $1^{\circ}C$. 
At each temperature the sample is kept for about $10$min to equilibrate and 
$g_2(t)$ is recorded over an additional $10$ min. 
For our microgel suspensions, $g_2(t)$ (partially) relaxes in a characteristic time $\tau$ that depends on temperature.
In the case of more dilute suspensions, $g_2(t)$ exhibits a roughly exponential decay characteristic of simple diffusion,
with $\left\langle {\Delta r^2 (t)} \right\rangle = 6 D_s(\phi) t$.  However, for concentrated suspensions, the behavior is more
complex.  At temperatures  between $27^{\circ}$C  and $30^{\circ}$C, $g_2(t)$ exhibits a slightly stretched exponential decay and a corresponding 
power-law MSD, $\left\langle {\Delta r^2 (t)} \right\rangle\sim t^{\beta}$ with $\beta\lsim1$, characteristic of a 
crowded suspension of soft, repulsive particles. 

\begin{figure}[h!]
\includegraphics[trim=0cm 2.6cm 0cm 0cm,clip,width=7cm]{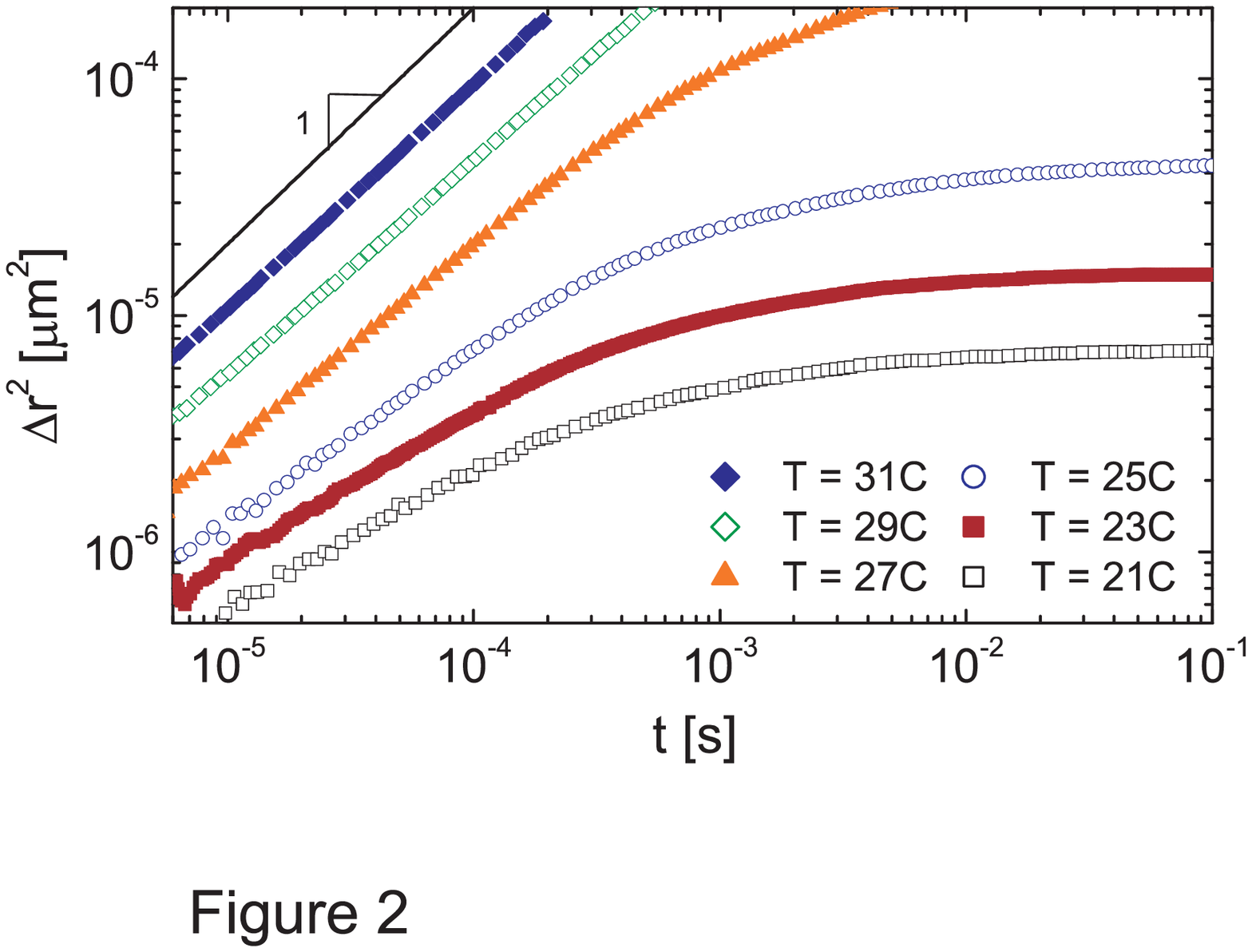}
\caption{ Thermally activated nanoscale motion. Mean square displacement $\left\langle {\Delta r^2 (t)} \right\rangle$ of the microgel particles from diffusing wave spectroscopy.}
\label{dwsmsd} 
\end{figure}
\begin{figure}[h!]
\includegraphics[trim=0cm .5cm 0cm 0.3cm,width=7.5cm]{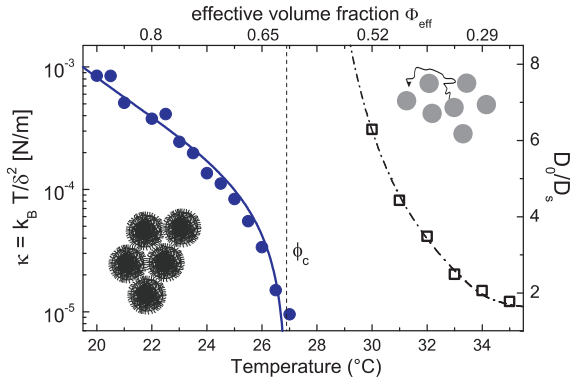}
\caption{Microgel liquid-solid transition as a function of temperature. The transition is driven by the change in effective
particle volume due to swelling. The reciprocal short-time diffusion coefficient $D_0/D_s$ (open squares) increases strongly
when approaching the transition. Dash-dotted line: semi emprical expression by Lionberger and Russel for hard spheres. 
Solid circles: elastic spring constant $\kappa=k_B T/\delta^2$ from DWS.  
Solid line is calculated from Eq.(\ref{Phieff}) with ($\Phi_c=0.625,\alpha=R_b/R=0.8$)} 
\label{springd} 
\end{figure}
\begin{figure}[h!]
\includegraphics[trim=0cm 2cm 0cm 0cm,clip,width=7cm]{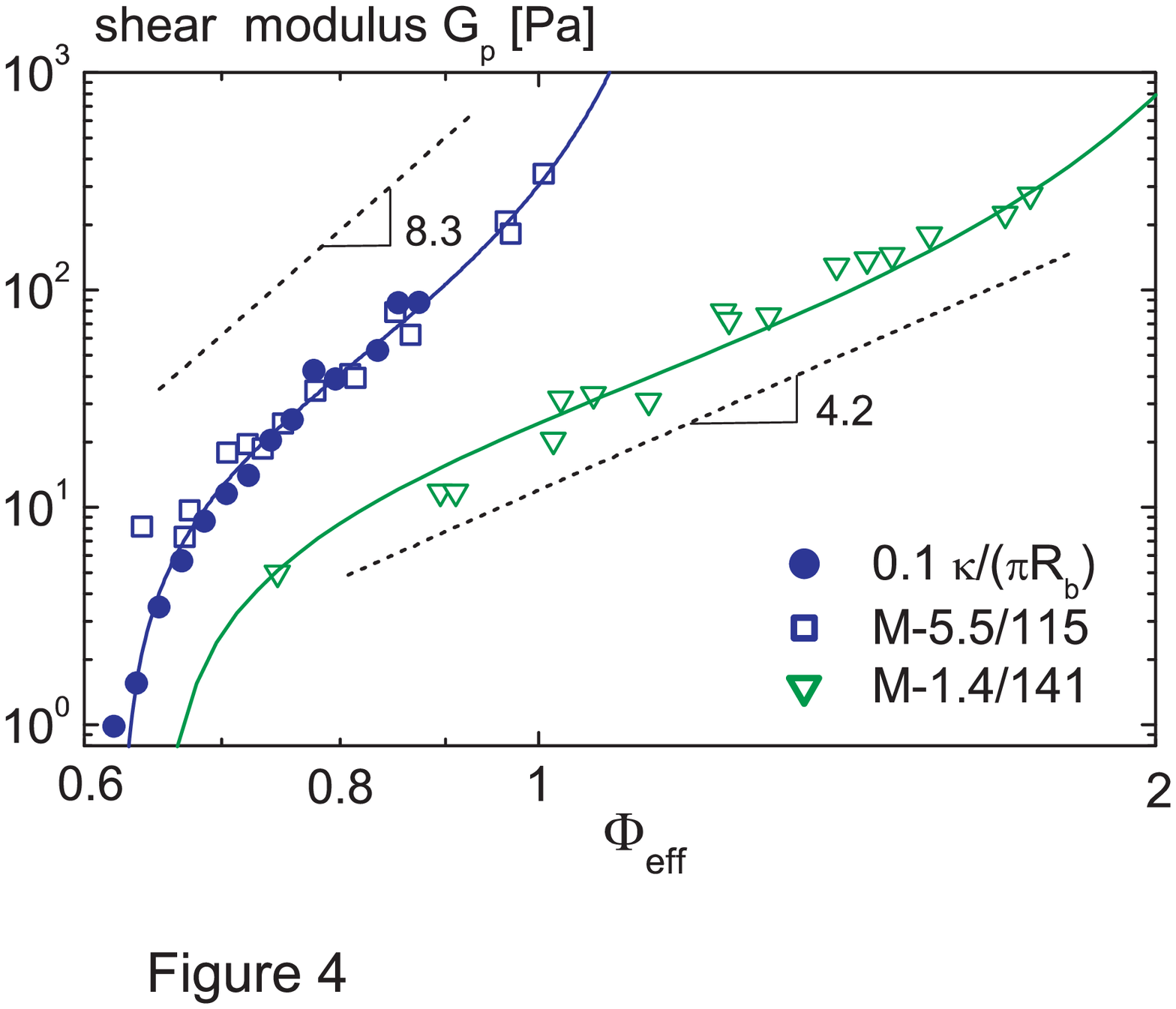}
\caption{Shear modulus of quenched microgels as a function packing density. Full symbols are derived from the diffusing wave spectroscopy 
data shown in Fig. \ref{dwsmsd} and \ref{springd}. Open symbols: data reproduced from Ref. \cite{Stieger2004}. 
Squares (M-5.5/115): cross-linker concentration 5.5 mol $\%$ BIS and hydrodynamic radius $R_H(25^{\circ}$C$)=115$nm. 
Open triangles (M-1.4/141): 1.4 mol $\%$ BIS and $R_H(25^{\circ}$C$)=141$nm. 
Solid lines are calculated from Eq.(\ref{Phieff}) with ($\Phi_c=0.625,\alpha=R_b/R=0.8$) and ($\Phi_c=0.65,\alpha=R_b/R=0.63$).  
Dashed straight lines indicate a power-law scaling $G_p \propto \Phi_{eff}^{m}$ with $m=1+n/3$. } 
\label{moduluscomparison} 
\end{figure}

\indent However, at $T=27^{\circ}$C and below the particles have swollen into contact and the
short-time decay of  $g_2(t)$ is only a partial one, leading to a plateau value of the MSD characteristic of the elastic 
properties of the jammed suspension \cite{glasstransition}. Figure \ref{dwsmsd}  shows plots of $\left\langle {\Delta r^2 (t)} \right\rangle$ 
versus time $t$ for a range of temperatures above and below the jamming transition.
\newline \indent We first analyze the influence of crowding on the temperature dependence of the self diffusion coefficient $D_s(\phi)$ for $T \ge 30^{\circ}$C. As shown in Fig. \ref{springd} our results are well described by the semiempirical expression of Lionberger and Russel 
for hard spheres \cite{Lion94}:  $ {{D_0 } \mathord{\left/
 {\vphantom {{D_0 } D_s(\phi)}} \right.
 \kern-\nulldelimiterspace} D_s(\Phi_{eff})} = \left[ {(1 - 1.56\,\Phi_{eff} )(1 - 0.27\,\Phi_{eff} )} \right]^{ - 1} $, 
 where $\Phi _{eff}  = \eta \times {{4\pi R^3 } \mathord{\left/
 {\vphantom {{4\pi R^3 } 3}} \right.
 \kern-\nulldelimiterspace} 3} $. Here we have introduced an effective steric radius $R=R_H \times 1.027$ that reflects quantitatively the hard-sphere like behavior in the liquid state.
\indent At temperatures $27^{\circ}$C and below, where the system has entered a viscoelastic solid phase,
the swollen particles are trapped and the assembly is dynamically arrested.  
We know that in the swollen state the particle's mass is primarily concentrated in a sphere of size $R_b$, 
whereas the dynamic properties are governed by the effective size $R$ which includes a compliant corona, 
$R \simeq R_b + L_0$. 
We can therefore consider the thermal motion of the particle core, radius $R_b$, as a probe of the particle interactions 
defined by the quenched particle coronas. The particle MSD shown in Fig. \ref{dwsmsd} clearly indicates a 
plateau value at long times.  From the equipartition of energy a local spring constant can be defined by the  maximum value $\delta^2$ of the MSD 
in the plateau region, $\kappa \simeq kT/\delta^2$,  
which can be directly related to the interaction potential $\psi$ between two spheres $\kappa = \partial ^2 \psi /\partial r^2$.   
In Fig. \ref{springd} we show the temperature dependence of $\kappa$ in the arrested state, determined from the mean square displacement at t $\sim 0.1$s. 
\newline \indent As outlined above we propose that the particle corona can be modeled as a polymer brush of thickness $L_0$~\cite{Alexander77,Milner88} that extends into the good solvent from a reference surface located at $R_b$. Since  typically $L_0 \ll R_b$ curvature effects on the brush configuration and the particle interaction potential should be small. Thus the force between two particles separated by $r=2R_b+d$ may be written in the Derjaguin approximation \cite{Derjaguin34,Likos2000} as

\begin{equation}
F(d) = \pi R_b \int\limits_{2L_0 }^d {f(x)dx} 
\end{equation}
where $f(d)$ is the force per unit area between two flat surfaces \cite{noteDerj}. The Alexander - de Gennes scaling model \cite{Alexander77}
predicts an $f(d)$ of the form
\begin{equation}
f(d) \sim k_B T/s^3 
 \left[ {(2L_0 /d)^{9/4}  - (d/2L_0 )^{3/4} } \right]
\label{brush}
\end{equation}
with an unknown prefactor of order 1. The first term in Eq \ref{brush} is set by the osmotic repulsion of a semidilute solution of equal density whereas the second term is the chain tension due to the entropic penalty upon stretching the grafted chains. The spring constant is thus $\kappa = \partial F(d)/\partial d =  \pi R_b f(d) $.
Defining a characteristic ratio $\alpha=R_b/R<1$
we can replace $d/2L_0$ by $ {{({r \mathord{\left/
 {\vphantom {r R}} \right.
 \kern-\nulldelimiterspace} R} - 2\alpha )} \mathord{\left/
 {\vphantom {{({r \mathord{\left/
 {\vphantom {r R}} \right.
 \kern-\nulldelimiterspace} R} - 2\alpha )} {(2 - 2\alpha }}} \right.
 \kern-\nulldelimiterspace} {(2 - 2\alpha }})$ \cite{note3}. 
We can now write the spring constant in terms of the reduced volume fraction $\tilde \Phi=\Phi_{eff}/\Phi_c$ 
 since $r/R \simeq 2 \tilde \Phi^{-1/3}$.  For random close packing conditions, $\Phi_{c} \simeq 0.64$.  
 Note that at initial contact, $r=2R$,  we recover the liquid-to-solid transition. 
 For simplicity we neglect the weak temperature dependence 
 of $R_b$ and find : 
\begin{equation}
\kappa  \propto \left( {\frac{{1 - \alpha }}{{\tilde \Phi^{ - 1/3}  - \alpha }}} \right)^{9/4}  - \left( {\frac{{\tilde \Phi^{ - 1/3}  - \alpha }}{{1 - \alpha }}} \right)^{3/4} 
 \label{Phieff}
\end{equation}
\indent The spring constant directly sets the scale for the macroscopic shear modulus via $\pi R G_p \sim \kappa$. 
This can be obtained at a scaling level by equating the displacement energy $\delta^2 R G_p$ with the thermal energy $kT$. 
For a multiparticle system, the same scaling relation can also be derived from the statistical mechanical theory 
of Zwanzig and Mountain \cite{Zwanzig65,Buscall91}. Figures  \ref{springd} and  \ref{moduluscomparison} show the excellent agreement of Eq. \ref{Phieff} with the DWS-data using 
$R_b/R = 0.8$ and $\Phi_c = 0.625$. To cover an even larger range of effective volume fractions we include in our analysis literature data of two well characterized microgel systems \cite{Stieger2004}.  The two systems are similar to ours except for the lower amount of BIS cross-linker used for one of the systems. Interestingly the moduli of the highly cross-linked particles and our data can be collapsed on single curve, whereas the particles with a lower amount of cross-linker are substantially softer.  Moreover, both the more rigid and the soft particle rheology 
can be described by our model.  Note that, over a limited range of densities, $G_p(\Phi_{eff})$ 
resembles a power-law with an exponent $m$ that depends strongly on $\alpha=R_b/R$. 
\newline \indent In this work we have focused our attention on the general scaling concepts that govern the linear elasticity in microgel pastes. We showed that a simple polymer brush model captures the essential physics of the interparticle interactions, regardless of the various approximations made. Moreover, we were able to make detailed predictions for the bulk elastic modulus over a large range of effective volume fractions. Our results explain the origin of the power-law scaling of the elastic modus in these systems, which emerges naturally as a consequence of interparticle interactions mediated by  brushlike coronas of
the microgel particle constituents.   This simple picture should have broad utility for understanding the elastic properties of 
this important class of disordered soft solids.

This work was supported by the Top-Nano 21 Project 5971.2, the Swiss National Science Foundation and Marie Curie network Grant No. MRTN-CT2003-504712. JLH acknowledges support from an NSERC discovery grant. The authors thank Fyl Pincus, Peter Schurtenberger, Mathias Fuchs, Luis Rojas, Frederic Cardinaux, Jerome Crassous and Veronique Trappe for interesting discussions.

\end{document}